\documentclass[12pt,preprint]{aastex}

\def\lapp{\ifmmode\stackrel{<}{_{\sim}}\else$\stackrel{<}{_{\sim}}$\fi}
\def\gapp{\ifmmode\stackrel{>}{_{\sim}}\else$\stackrel{>}{_{\sim}}$\fi}

\newcommand{\xte}{\textit{RXTE}}

\newcommand{\tfe}{1E~1048.1--5937}
\newcommand{\tfn}{1E~2259.1+586}
\newcommand{\soelong}{RXS~J170849.0--400910}
\newcommand{\soe}{1RXS~1708--4009}
\newcommand{\oft}{4U~0142+61}
\newcommand{\efo}{1E~1841--045}
\newcommand{\axj}{AX~J1845.0--0258}
\newcommand{\unitfrac}[2]{\frac{\textrm{#1}}{\textrm{#2}}}
\newcommand{\nbins}{N_{\phi}}

\newcommand{\factor}{\mathcal{F}}
\newcommand{\DC}{\mathcal{DC}}

\begin{document}

\title{Long-Term \xte\ Monitoring of Anomalous X-ray Pulsars}

\author{ 
Fotis P. Gavriil\altaffilmark{1} , Victoria M. Kaspi \altaffilmark{1,2,3}}

\altaffiltext{1}{Department of Physics, Rutherford Physics Building,
McGill University, 3600 University Street, Montreal, Quebec,
H3A 2T8, Canada}

\altaffiltext{2}{Department of Physics and Center for Space Research,
Massachusetts Institute of Technology, Cambridge, MA 02139}

\altaffiltext{3}{Alfred P. Sloan Research Fellow}

\begin{abstract}

We report on long-term monitoring of three anomalous X-ray pulsars
using the \textit{Rossi X-ray Timing Explorer} ({\it RXTE}). We present
a phase-coherent timing ephemeris for \oft, and show that it has
rotated with high stability over 4.4~yr, with RMS phase deviations of
7\% of the pulse period from a simple fit including only $\nu$ and
$\dot{\nu}$.  We report on the continued timing stability of \tfn, for
which phase coherence has now been maintained over 4.5~yr, as well as
on the detection of a significant $\ddot{\nu}$ in 1.4~yr of monitoring
of \soelong, consistent with recovery following a glitch.  We note a
correlation in which timing stability in AXPs decreases with increasing
$\dot{\nu}$.  The timing stability of soft gamma repeaters in
quiescence is consistent with this trend, given their large spin-down
rates.  This trend is similar to one seen in radio pulsars, suggesting
a connection between the three populations.  We find no large
variability in pulse morphology as a function of time.  We present high
signal-to-noise ratio average pulse profiles for each AXP, and consider
them as a function of energy.  We find a variety of different
behaviors, and consider possible trends in the data.  We also find no
large variations in pulsed flux, and set $1\sigma$ upper limits of
$\sim$20--30\% of the mean.
\end{abstract}


\section{INTRODUCTION}

Anomalous X-ray pulsars are an unusual class of astrophysical objects. 
There are currently only five confirmed AXPs: \oft,
\tfe, \efo, \soelong\ (hereafter \soe), and \tfn.  All five are
located in the Galactic plane and two (\tfn\ and \efo) are coincident
with supernova remnants.  A sixth AXP candidate, \axj, is also
coincident with a supernova remnant \citep{ggv99}.

The observational properties of AXPs \citep[see][for a review]{ims02}
are generally quite different from those of conventional accreting binary
X-ray pulsars \citep[see][and references therein for a compendium of
accreting X-ray pulsar properties]{bcc+97}.  In particular, AXPs have
spin periods in the narrow 6--12~s range, while accreting X-ray pulsars
have periods that span 0.002--1400~s.  AXPs have luminosities in the
narrow range $10^{34}-10^{36}$~erg~s$^{-1}$.  By comparison, typical
accreting X-ray pulsars are extremely variable, with luminosities
spanning a much broader range ($10^{34}-10^{39}$~erg~s$^{-1}$).  AXP
spectra are considerably softer than those of the typical accreting
pulsar.  Accreting pulsar spectra are generally well described as a
power law with a cut-off in the 10--20~keV range.  AXP spectra, by
contrast, are generally characterized by a blackbody spectrum of
energy $kT \simeq 0.35-0.6$~keV, with a hard energy excess that can be
characterized by a steep power law of index $-2$ to $-4$.  The timing
properties of AXPs also contrast sharply with those of most accreting
pulsars.  All of the known AXPs, as we show in this paper, appear to be
undergoing steady, prolonged spin-down, though some are more steady
than others.  By comparison, most accreting pulsars are spinning up (on
average), or alternating between spin-up and spin-down, and have poor
rotational stability.

Most importantly, the AXPs show no evidence for a binary companion.  Specifically, the limits on
the X-ray/optical flux ratio in AXPs rule out the presence of massive
companions as in the conventional high-mass X-ray binaries
\citep[e.g.][]{mcb92,cm97,ics+99}.  In addition, careful pulse timing of AXPs
has failed to detect any evidence for binary motion of the neutron star
on time scales of a few minutes to several days \citep{mis98,wdf+99}.
The limits imply that only extremely low-mass companions are allowed in
these systems, unless we are observing all AXPs face-on, which is unlikely.
Furthermore, mass-transferring low-mass companions are also 
unlikely because the supernova remnant associations imply
youth, and because  these systems rarely  survive the supernova.

Currently, models for AXPs fall into two distinct categories.  One class of 
models is that AXPs are young, isolated, ultra-magnetized
neutron stars or ``magnetars'' \citep{td96a,hh97}. If the spin-down of
the pulsar is primarily due to magnetic dipole radiation,
then the AXPs have enormous surface magnetic
dipolar fields, in the range $B\sim10^{14}-10^{15}$~G.  The
identification with magnetars is further strongly motivated by the
similarity of the AXP emission to that of the soft gamma-ray repeaters
(SGRs) in quiescence.  Specifically, the latter have similar pulse
periods, are spinning down \citep{kds+98,ksh+99}, and have X-ray
spectra that are comparable to, though somewhat harder than, those of
the AXPs, at least when not in outburst \citep{mcft00,kkm+01}.
Independent evidence for the ultra-high magnetic fields exists in SGRs;
for example, a $\sim 10^{15}$~G magnetic field is required to contain
the radiation that is seen following major outbursts
\citep{td95,td96a}.

The second category of model proposed to explain AXP emission is that
they are accreting from a disk of material leftover from the supernova
explosion that created the neutron star \citep{chn00,alp99,mlrh01}.
Fall-back disks around neutron stars have been proposed in other
contexts \citep{mic88}; here, the AXPs represent a single, short-lived
phase in the evolution of a neutron star having magnetic field only
slightly larger than those of radio pulsars.  In this phase, the
neutron star is in a propeller mode in which the inner edge of
the accretion radius lies between the magnetospheric and corotation
radii.  This model, in its original version \citep{phn00},
significantly overpredicted the expected optical/IR flux from the
fall-back disk \citep{hkvk00,hvk00,kkk+01}.  A more recent version
includes the effects of a thermal disk instability which may prevent
the disk from expanding radially, hence allowing it to remain
sufficiently dim to be consistent with the optical/IR limits
\citep{mrl01}.  Nevertheless, difficulties remain with the accretion
model.  In particular, although the noise properties of a fall-back
disk are currently unknown, the extremely steady spin down of two AXPs
(Kaspi, Chakrabarty \& Steinberger 1999) \nocite{kcs99} is at odds with
that seen in most X-ray pulsars that definitely have accretion disks
\citep[but see ][]{bia01}.  Furthermore, very noisy timing behavior
seen in one AXP (\tfe) shows no evidence for correlated flux
variations, as are expected in fall-back disk accretion models
\citep{kgc+01}.

In this paper, we present a variety of observational results from our
program of regular monitoring observations of AXPs with the {\it Rossi
X-ray Timing Explorer.} We make use of dedicated as well as archival
monitoring observations spanning several years for three AXPs.  Initial
results from this program were presented elsewhere (Kaspi et al. 1999,
Kaspi, Lackey \& Chakrabarty 2000) \nocite{klc00} 
and focused on the timing properties of \soe\ and \tfn.  \citet{kgc+01}
and \citet{ggk+01} present results for AXPs \tfe\ and \efo,
respectively.  In this paper, we report for the first time on \oft, and
update timing results for \soe\ and \tfn.  In addition, we use the same
data sets to study pulse morphology variability and energy dependence,
as well as the stability of the pulsed flux.  Our goal is to clarify
important observational properties of AXPs in the hope of offering
new, quantitative data for use in testing AXP models.

\section{OBSERVATIONS}
\label{sec:observations}

The results presented here were obtained using the Proportional Counter Array \citep[PCA;][]{jsg+96} on board the Rossi X-ray Timing Explorer (\xte ). The PCA consists of an array of five collimated xenon/methane multi-anode proportional counter units operating in the 2~--~60~keV range, with a total effective area of approximately $\rm{6500~cm^2}$ and a field of view of $\rm{\sim 1^o}$~FWHM. Our observations consist primarily of short snapshots  taken on a monthly basis (see Table~\ref{ta:exposures}). In addition, we used a handful of archival observations; the exposures in these observations vary from  very short ($T\simeq1$~ks) to very long ($T\simeq20-40$~ks; see Table~\ref{ta:exposures}). We used the \texttt{GoodXenonwithPropane} data mode, which records photon arrival times with 1-$\mu$s resolution and bins energies into one of 256 channels. Due to the soft spectrum of these sources, we analyzed only events from the top xenon layer of each PCU. Photon arrival times at each epoch were adjusted to the solar system barycenter, and binned with 31.25-ms time resolution.

\section{ANALYSIS AND RESULTS}
\label{sec:analysis and results}

\subsection{Phase Coherent Timing}
\label{sec:timing}

\begin{figure}
\plotone{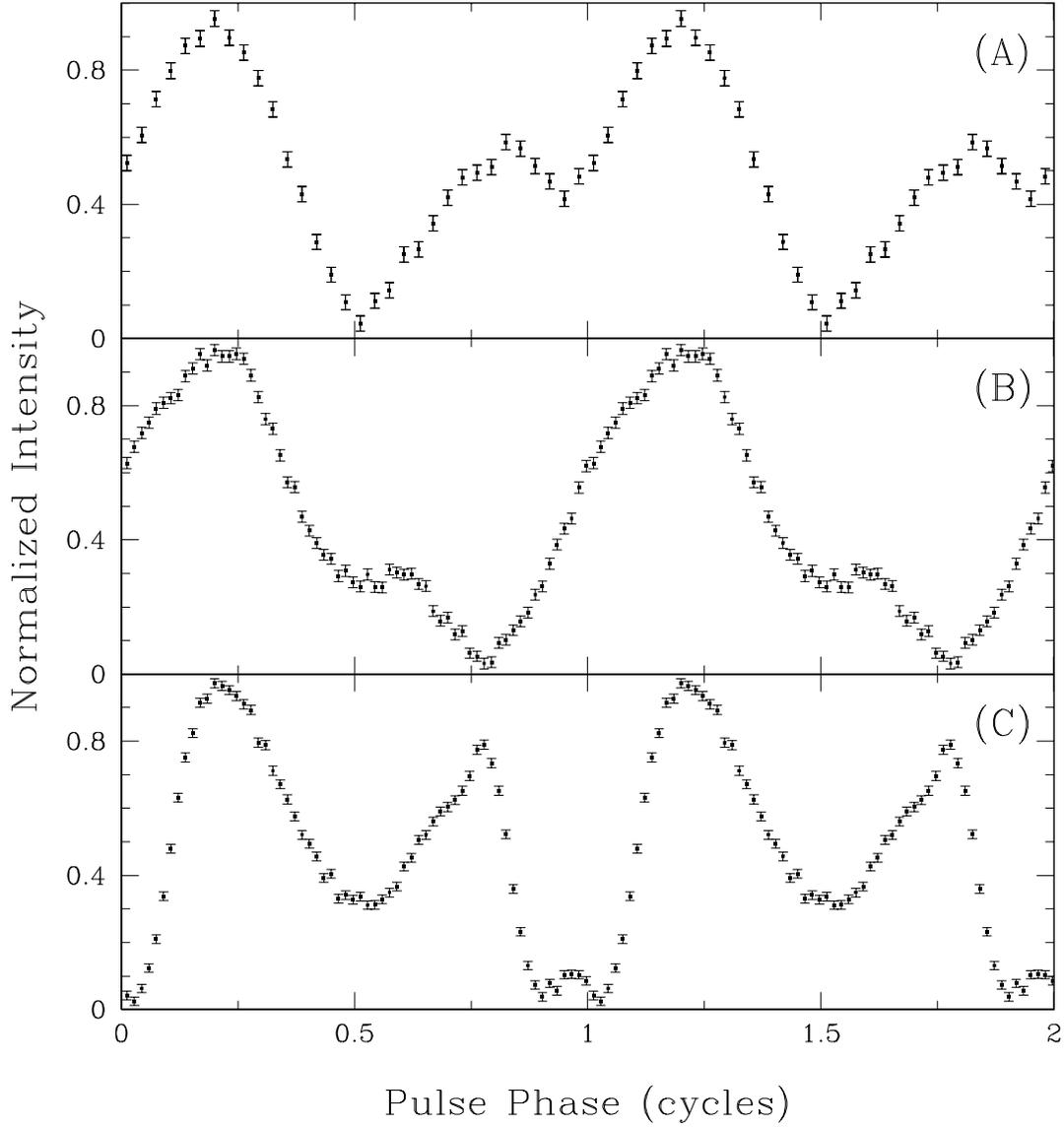}
\figcaption{Average pulse profiles of the AXPs. Two cyles are plotted for clarity. (A)~\oft\ (Total exposure time: 149~ks, energy range: 2.5~--~9.0~keV); (B)~\soe\ (185~ks, 2.5~--~9.0~keV); (C)~\tfn\ (469~ks, 2.2~--~5.5~keV).
\label{fig:templates}}
\end{figure}

\begin{figure}
\plotone{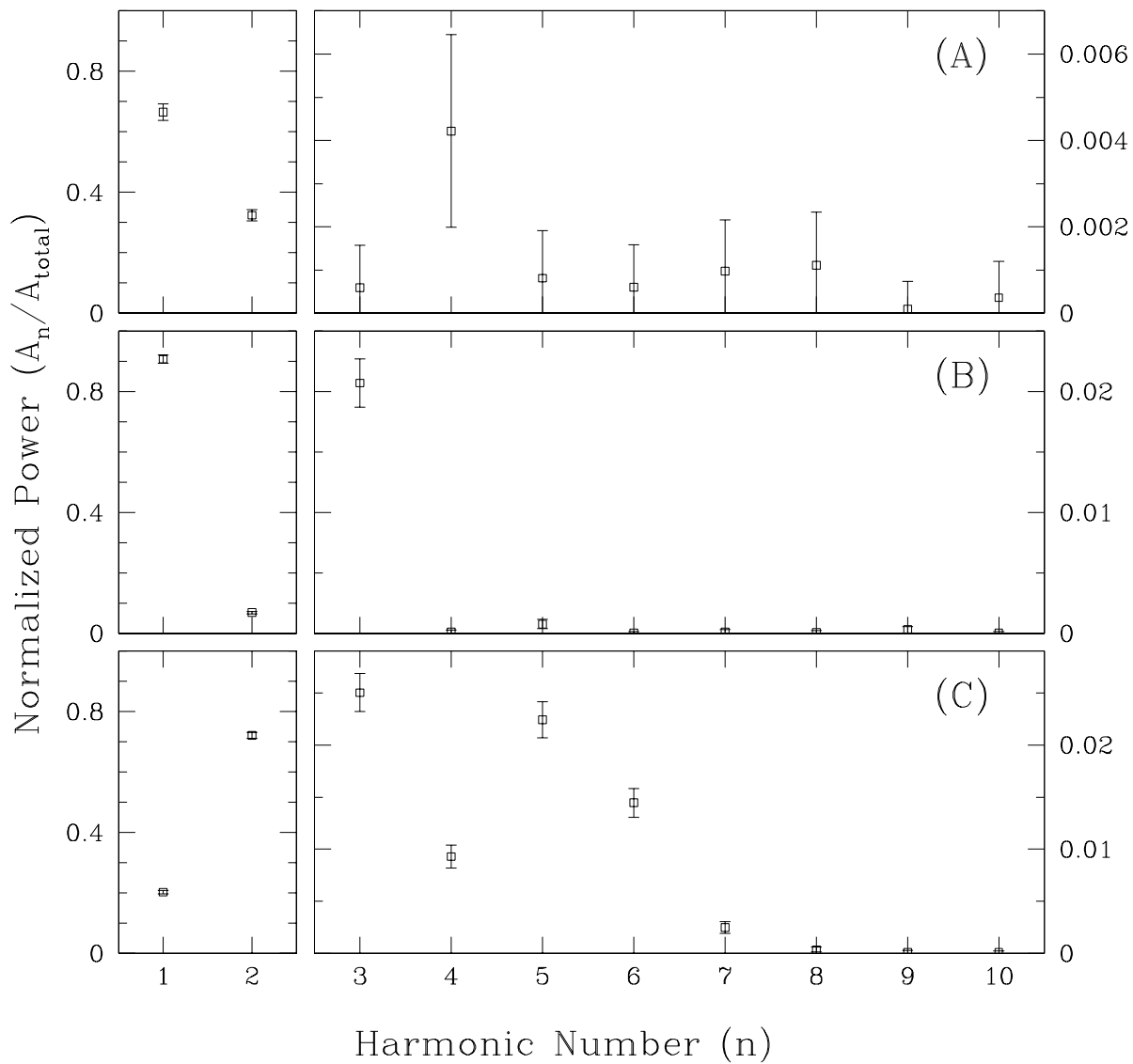}
\figcaption{Harmonic content of the average pulse profiles in Figure~\ref{fig:templates}. (A)~\oft\ (Energy range: 2.5~--~9.0~keV) ; (B)~\soe\ (2.5~--~9.0~keV); (C)~\tfn\ (2.2~--~5.5~keV). The ratio of the power in the $n^{\textrm{th}}$ harmonic to the total power in all harmonics is plotted versus  $n$.
\label{fig:fft_templates}}
\end{figure}

In the timing analysis, we included only those events having energies in a predetermined range so as to maximize the signal-to-noise ratio of the pulse. The energy cuts used were 2.5~--~9.0~keV for \oft\ and \tfn\ and 2.2~--~5.5~keV for \soe.  Each binned time series was epoch-folded using the best estimate frequency determined initially from either a periodogram or Fourier transform (though later folding was done using the timing ephemeris determined by maintaining phase coherence; see below). Resulting pulse profiles were cross-correlated in the Fourier domain with a high signal-to-noise template created by adding phase-aligned profiles from previous observations. The templates are shown in Figure~\ref{fig:templates} along with their respective Fourier transforms in Figure~\ref{fig:fft_templates}.  The cross-correlation returns an average pulse time-of-arrival (TOA) for each observation corresponding to a fixed pulse phase. The pulse phase $\phi$ at any time $t$ can be expressed as a Taylor expansion,

\begin{equation}
\phi(t) = \phi(t_0) + \nu_0 (t-t_0) + \frac{1}{2} \dot{\nu}_0(t-t_0)^2   + \frac{1}{6} \ddot{\nu}_0(t-t_0)^3 + \cdots,
\label{eq:phase}
\end{equation}

\noindent where $\nu\equiv 1/P$ is the pulse frequency,  $\dot{\nu} \equiv d\nu/dt$, etc., and subscript `$0$' denotes a parameter evaluated at the reference epoch $t=t_0$. The TOAs were fit to the above polynomial using the pulsar timing software package \texttt{TEMPO}\footnote{http://pulsar.princeton.edu/tempo}. Unambiguous pulse numbering is made possible by obtaining monitoring observations spaced so that the best-fit model parameters have a small enough uncertainty to allow prediction of the phase of the next observation to within $\sim0.2$. Typically this requires two closely spaced observations  (within a few of hours of each other) followed by one spaced a few days later, and regular monitoring thereafter, as long as phase coherence can be maintained.

To minimize use of telescope time while maximizing precision, our
monitoring data consisted of frequent brief snapshots of each pulsar (Table~\ref{ta:exposures}). These snapshots suffice to measure TOAs to good precision.  However, at each epoch, the period as determined by a Fourier Transform or epoch-folding has typical uncertainty of a few milliseconds. Thus, snapshots can determine spin parameters with high precision only when phase coherence can be maintained. 

\begin{figure}
\plotone{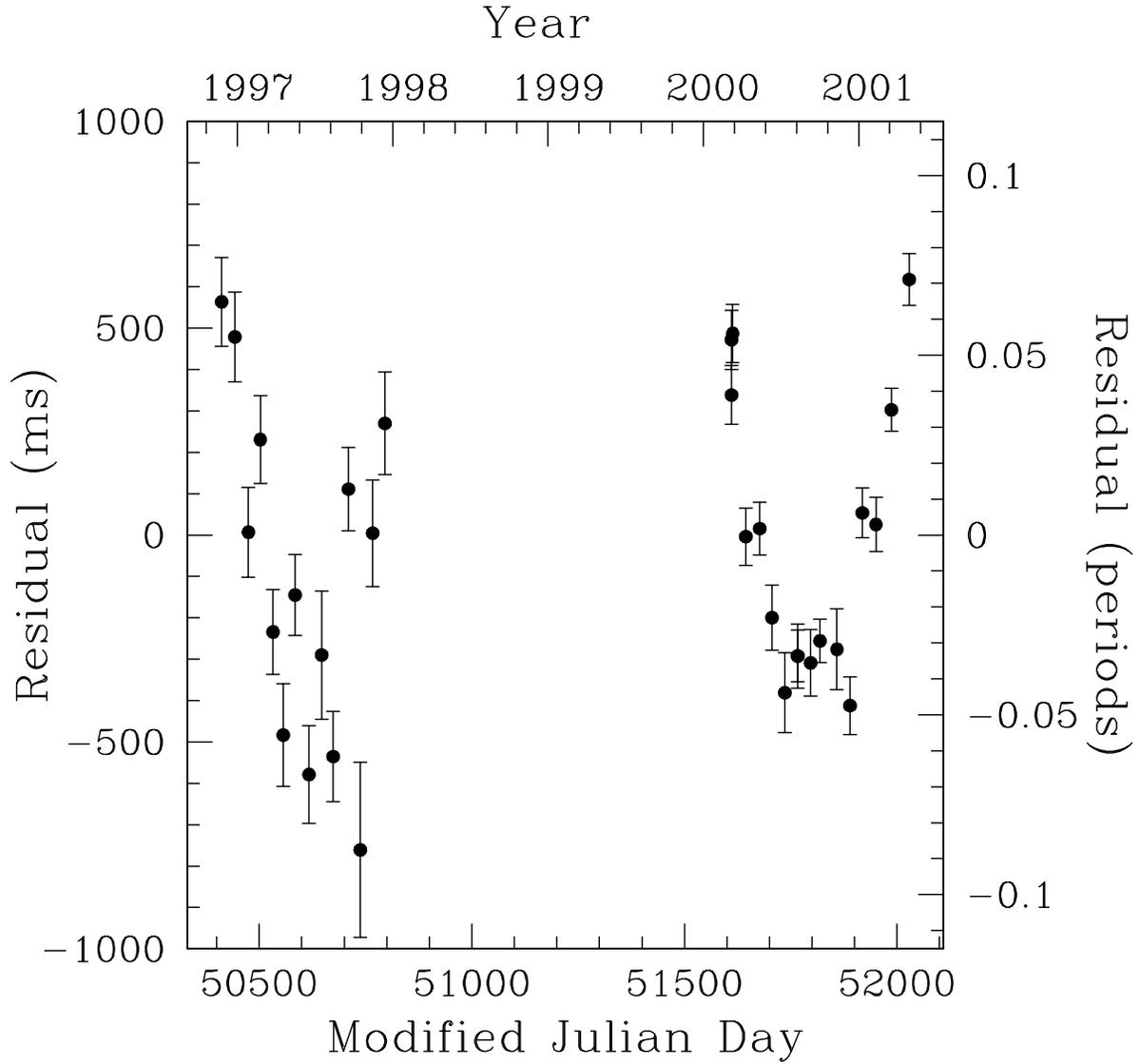}
\figcaption{Arrival time residuals for \oft\ with Ephemeris A (Table~\ref{ta:timing0142}), that includes $\nu$ and $\dot{\nu}$, subtracted.  Residuals relative to Ephemeris B are qualitatively similar.
\label{fig:res0142}}
\end{figure}

For \oft, we report here for the first time a phase-coherent timing
solution that indicates that this AXP has been an extremely stable
rotator over 4.4~yr of {\it RXTE} monitoring.  However, a 2-yr gap in
the data precludes unambiguous absolute pulse numbering; we find two
equally viable phase-connections that differ by one pulse.  Unambiguous
phase count can be maintained in 1.3 and 1.4-yr segments at the
beginning and end of the 4.4-yr span, respectively.  Phase residuals
after subtraction of spin-down models that include $\nu$ and
$\dot{\nu}$ only in the first and second segments have RMS of 2.7\% and 1\% of
the pulse period, respectively, and are featureless.  The best-fit
model parameters for each segment, as well as the two indistinguishable
models for the entire 4.4-yr span, are given in
Table~\ref{ta:timing0142}.  Timing residuals for solution ``A'' (see
Table~\ref{ta:timing0142}) are shown in Figure~\ref{fig:res0142}; those
for solution ``B'' are qualitatively similar.  

\begin{figure}
\plotone{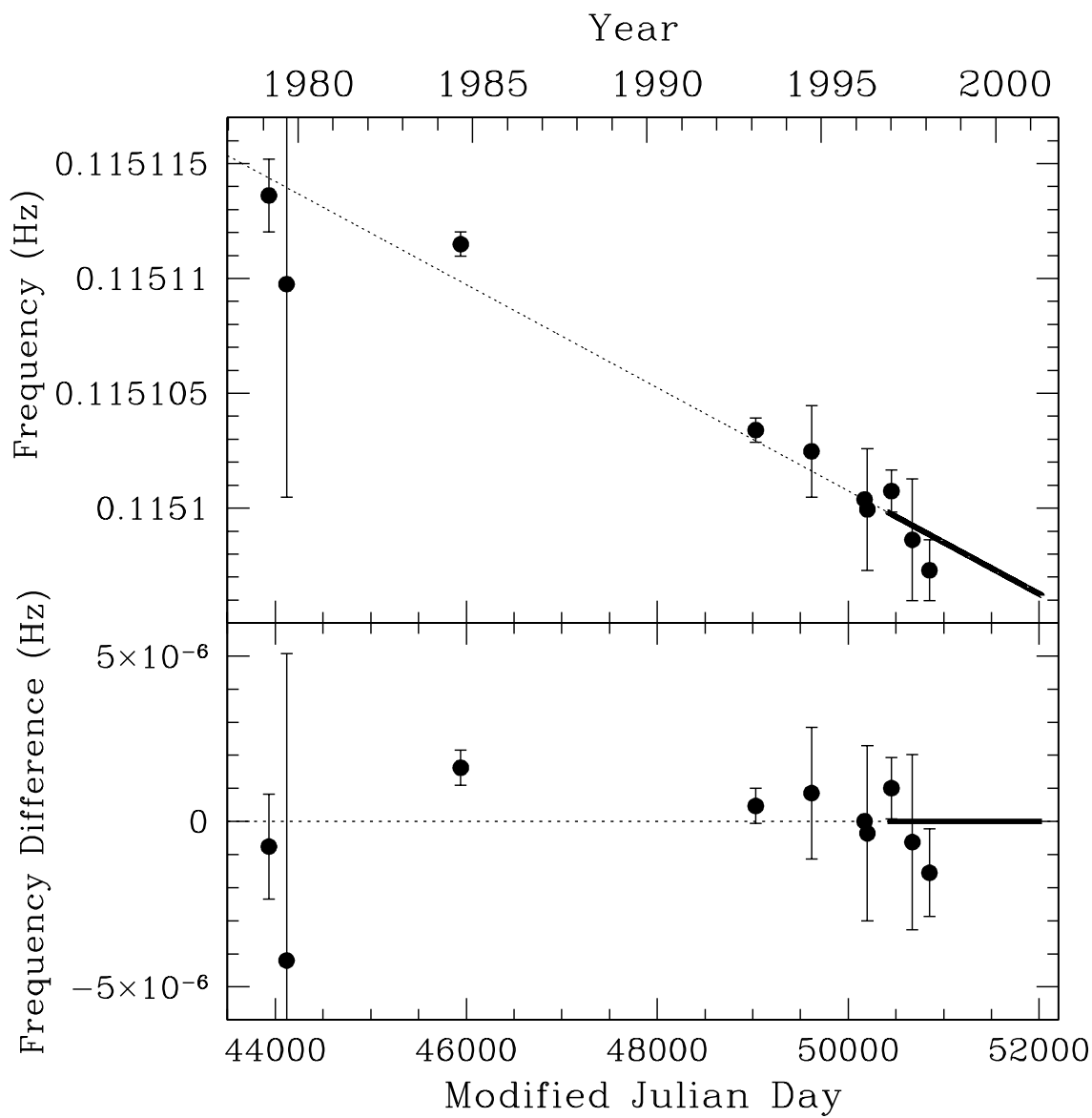}
\figcaption{Pulse frequency history for \oft.  Top panel: the solid, thick line is the ephemeris detemined from our phase-coherent timing with {\it RXTE}.  Individual data points are from a variety of X-ray missions (see \citeauthor{ioa+99} \citeyear{ioa+99} for references) and are shown with $90\%$ confidence level error bars.  Note that on this scale, Ephemeris A and B (see \S~\ref{sec:timing}, Table~\ref{ta:timing0142}) are indistinguishable.  The dotted line is the extrapolation of our fit backward.  Bottom panel:  the same data but with the linear trend and offset removed.  In both plots, the uncertainties (including that due to the ambiguity between A and B) are smaller than the width of the lines.
\label{fig:freq0142}}
\end{figure}

It is interesting to
compare our results with frequency measurements made for this pulsar
with various X-ray telescopes over the past $\sim$20~yr.
Figure~\ref{fig:freq0142} shows these data (taken from
\citeauthor{ioa+99} \citeyear{ioa+99}) along with our {\it RXTE}
ephemeris.  The top panel shows $\nu$ versus time; the thick solid line
is our phase-connected ephemeris (the distinction between A and B is
not visible on this plot) over the 4.4-yr span {\it RXTE} has monitored
it, while the thin line is the extrapolation of our ephemeris
backward.  The uncertainties are smaller than the thickness of the
lines.  The bottom panel shows the same data but with the linear and
offset trend removed.  All but one data point (taken with EXOSAT) agree
with the back-extrapolation.  That one point suggests that the pulsar
may exhibit deviations from a simple spin-down law; these could be due to random noise processes or glitches. Further observations can test this.

\begin{figure}
\plotone{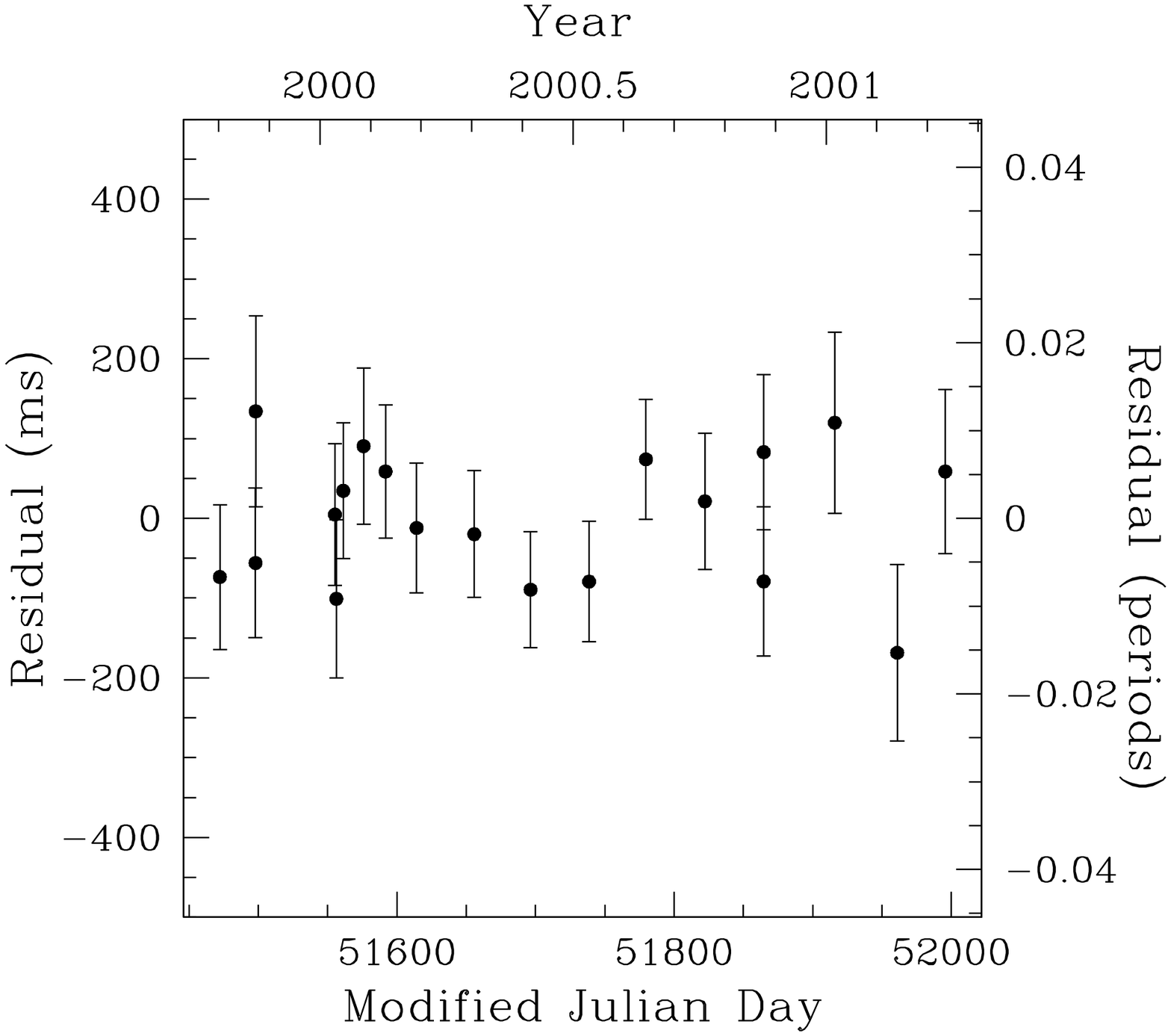}
\figcaption{Arrival time residuals for \soe\ with $\nu$, $\dot{\nu}$ and $\ddot{\nu}$ subtracted. 
\label{fig:res1708}}
\end{figure}

For \soe, phase coherent timing has been accomplished in the 1.4~yr
since the glitch reported by \citet{klc00}. In these data, we find a
significant positive $\ddot{\nu}$ (see Table~\ref{ta:timing}). This
indicates a decay of the negative $\dot{\nu}$, expected for long-term
glitch recovery, as seen in glitching radio pulsars
\citep[e.g.][]{sl96}. Phase residuals after subtraction of the best fit
model are shown in Figure~\ref{fig:res1708}. We note that the value for
$\dot{\nu}$ reported in Table~2 is different by $3.8\sigma$ from that
reported by \citet{klc00}.  This is likely due to the unmodeled
influence of $\ddot{\nu}$ in the earlier, shorter data set.

The rotational stability of \tfn, first reported by \cite{kcs99}, has
now persisted over 4.5~yr, although the inclusion of $\ddot{\nu}$ has
been necessary.  Phase residuals after subtraction of a simple model
that includes only $\nu$, $\dot{\nu}$ and $\ddot{\nu}$ have RMS under
1\% of the pulse period (see Table~\ref{ta:timing},
Fig.~\ref{fig:res1708}).  We note the presence of very low-level
systematic trends in the most recent data; their origin is unknown.

\subsection{Pulse Morphology Changes}
\label{sec:morph}

To search for pulse profile changes, the profiles were first phase aligned using the templates and the same cross-correlation procedure used for timing. Each data profile ($D$) was fit to the template ($T$) by adjusting two parameters, $\factor$ and $\DC$, to minimize a $\chi^2$ statistic, where 

\begin{equation}
\chi_{\nu}^2=\frac{1}{\nbins-4}\sum_{i=1}^{\nbins} { \frac{{\left(D_i - (\factor \cdot T_i - \DC)\right)}^2}{\sigma_{D_i}^2 + \factor^2\cdot\sigma_{T_i}^2 }}.
\label{eq:chi}
\end{equation}

\noindent Here $\nbins$ is the number phase bins, and $\sigma_D$ and
$\sigma_T$ are the errors associated with $D$ and $T$, respectively.
The resulting data profile was subtracted from the template to yield
``profile residuals'' for each observation.

To search for pulse profile variations, using the optimal $\factor$ and
$\DC$, we calculated the probability of each $\chi_{\nu}^2$ statistic
given their expected distribution and the number of degrees of freedom
($\nu = \nbins-4$).  The procedure was repeated for each pulsar with
$\nbins = 8, 16, 32$ in order to have sensitivity to a variety of types
of pulse profile changes.  We have not detected any large pulse profile
variations. This justifies our other analysis procedures which assume a
fixed profile.  However, in a handful of observations, we have found
evidence for low-level pulse profile changes, at the $\sim 3\sigma$
level for all three sources.  These are intriguing; however at this level, they require
confirmation using longer monitoring observations.

It is difficult to set quantitative upper limits on the amplitude of pulse profile
changes to which we were sensitive, as these depend on the shape of the change,
and vary depending on the length of the observation.  Typically, RMS profile
residuals are $\sim$20\% of the pulse peak, although this varied from 4--40\%.

\citet{ikh92} reported a significant change in the pulse morphology of
\tfn\ in 1.2--14~keV {\it GINGA} observations obtained in 1990, such that the
leading pulse had amplitude roughly half that of the trailing pulse
(compare with Fig.~\ref{fig:templates}C).  If correct, this has important
implications for the magnetar model, which only predicts such pulse
morphology changes in the event of a restructuring of the magnetic
field, as might occur following a major SGR-like outburst.
Although it is difficult to quantify upper limits on the amplitude of
pulse profile changes generally, we have done Monte Carlo simulations 
to see whether we were sensitive to the same change observed by \citet{ikh92}.
In particular, using the original 1990 {\it GINGA} pulse profile (kindly supplied to
us by B. Paul and F. Nagase) with random Poisson-deviated noise added, we repeated
our pulse change morphology analysis as described above.  Our simulations show that
we would have detected the change with high ($>$99\%) confidence in 25\% of our
observations (those with integration times $\gapp$10~ks), and with
moderate ($>$90\%) confidence in 45\% of our observations (those with integration
times of $\gapp$7~ks; see Table~\ref{ta:exposures}).

\subsection{Pulse Profile Energy Dependence}
\label{sec:energy}
AXP spectra are generally best  fit by a two-component model consisting of a 
photoelectrically absorbed blackbody with a hard power-law tail \citep{ioa+99}. Whether 
these two components are physically distinct is an open question . To investigate this, we
compared the pulse profile morphology in two energy bands, 2--4~keV and 6--8~keV. These 
were chosen because the 2--4~keV band has a significant blackbody  component ($\sim 40\%-60\%$ of total flux, depending on source and model) and the 6--8~keV band is greatly dominated by the power-law component ($\gapp 95\%$~of total flux). \citep[see][]{opk01}

\begin{figure}
\plotone{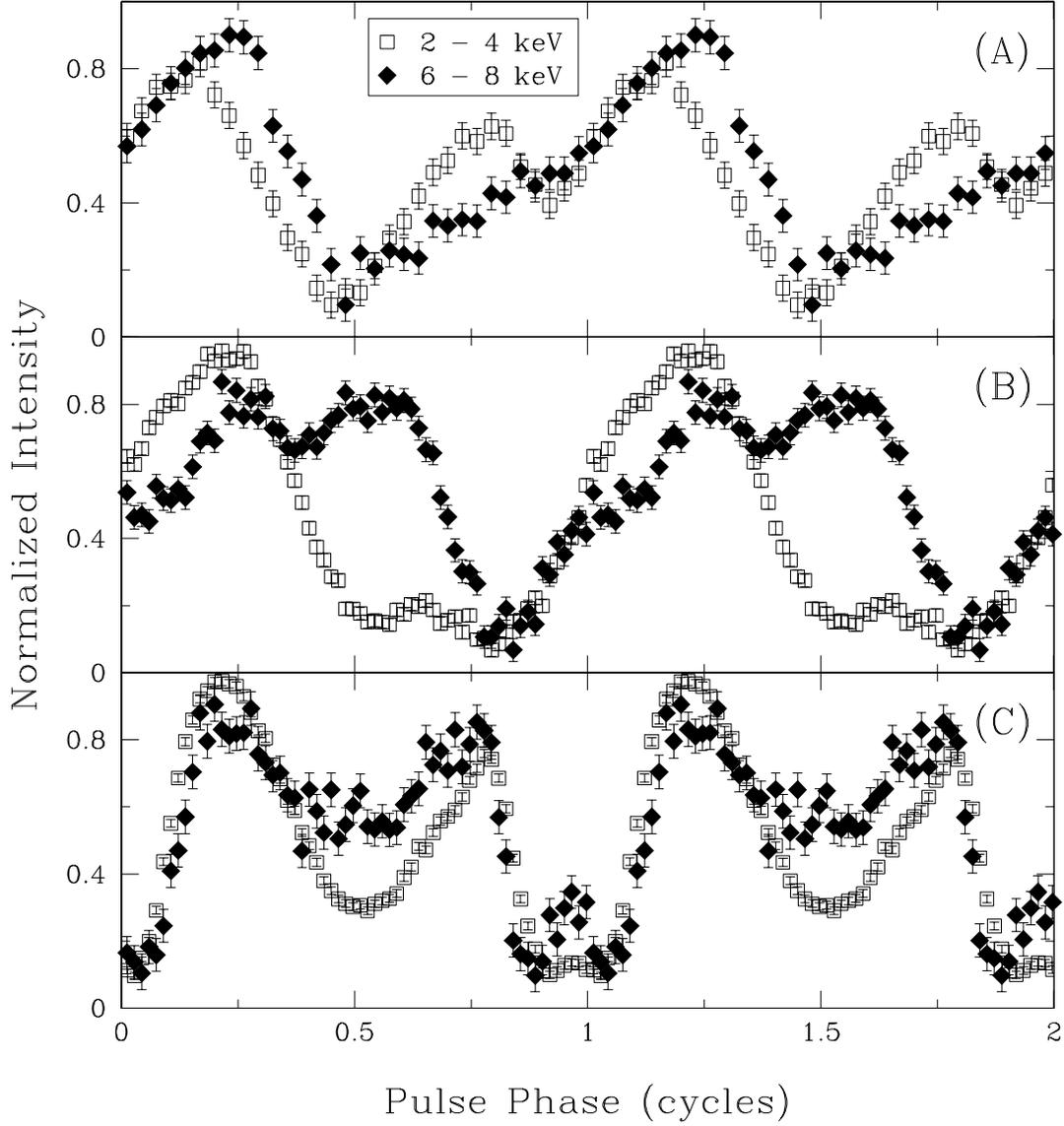}
\figcaption{Average pulse profile of (A)~\oft, (B)~\soe\ and (C)~\tfn\ in two energy bands as observed by \xte. Two cycles are plotted for clarity. Note that the scaling was chosen  to minimize the $\chi^2$ of the  difference between the two profiles.  Thus the only information 
that these plots convey is the relative amplitudes of the features of the profile.
\label{fig:energy}}
\end{figure}

Figure~\ref{fig:energy} display the average pulse profiles of \oft, 
\soe\ and \tfn\ in the energy bands 2--4~keV and 6--8~keV. Note that the scaling in these plots was chosen to 
minimize the $\chi^2$ of the difference between the soft and hard profiles. Thus the only information 
that these plots convey is the relative amplitudes of the features of the profile. 
The minimization of the $\chi^2$ statistic (Eq.~\ref{eq:chi}) between profiles in the two energy bands 
was done in the same manner as for the pulse morphology changes (see  
\S\ref{sec:morph}).

Another way of characterizing the pulse morphology as function of energy is to consider the pulse 
profile harmonic content in different energy bands. Figure~\ref{fig:fft_energy} displays
 the power of the $n^{\textrm{th}}$ harmonic ($A_n$) in units of total power in the remaining harmonics 
($A_{total} = A_1 + A_2 +\cdots + A_{\frac{\nbins}{2}}$, where $\nbins$ is the number of phase bins), 
versus harmonic number $n$, for each source. 

From Figure~\ref{fig:energy}~and~\ref{fig:fft_energy} it is clear that all three of the AXP pulse profiles vary significantly as a function of energy, but to different degrees. 
\subsection{Pulsed Flux Time series}
\label{sec:flux}

Given the large field-of-view of the PCA, the low count rates for the sources relative to the background, and the fact that, for example, \tfn\ is in a supernova remnant, total flux measurements are difficult with our \xte\ data. Instead, we have determined the pulsed component of the flux, by using the off-pulse emission as a background estimator.
 
\begin{figure}
\plotone{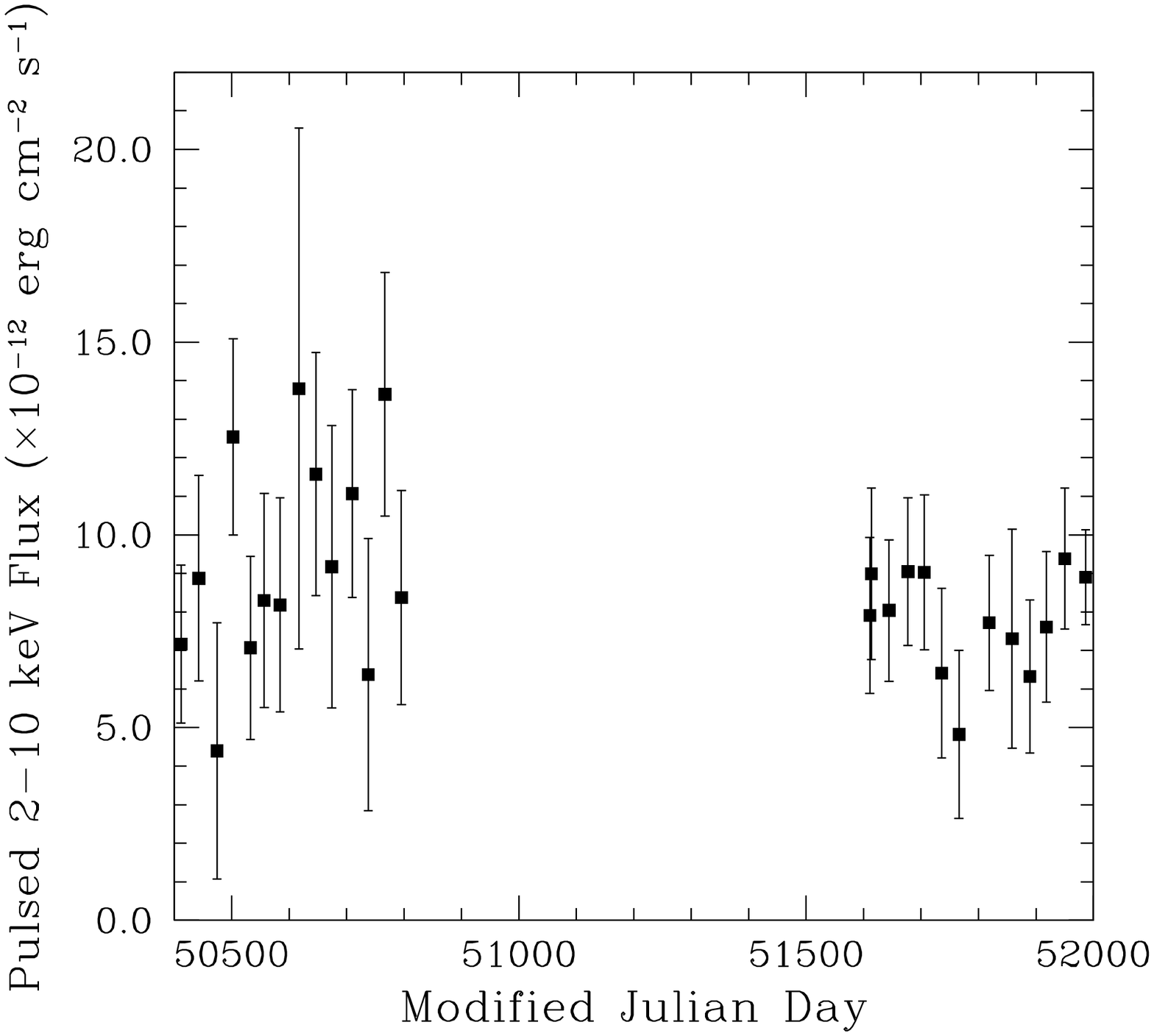}
\figcaption{Pulsed flux time series for \xte\ observations of \oft. Error bars represent $1\sigma$ confidence intervals. See \S\ref{sec:flux} for details of the analysis procedure.
\label{fig:flux0142}}
\end{figure}

\begin{figure}
\plotone{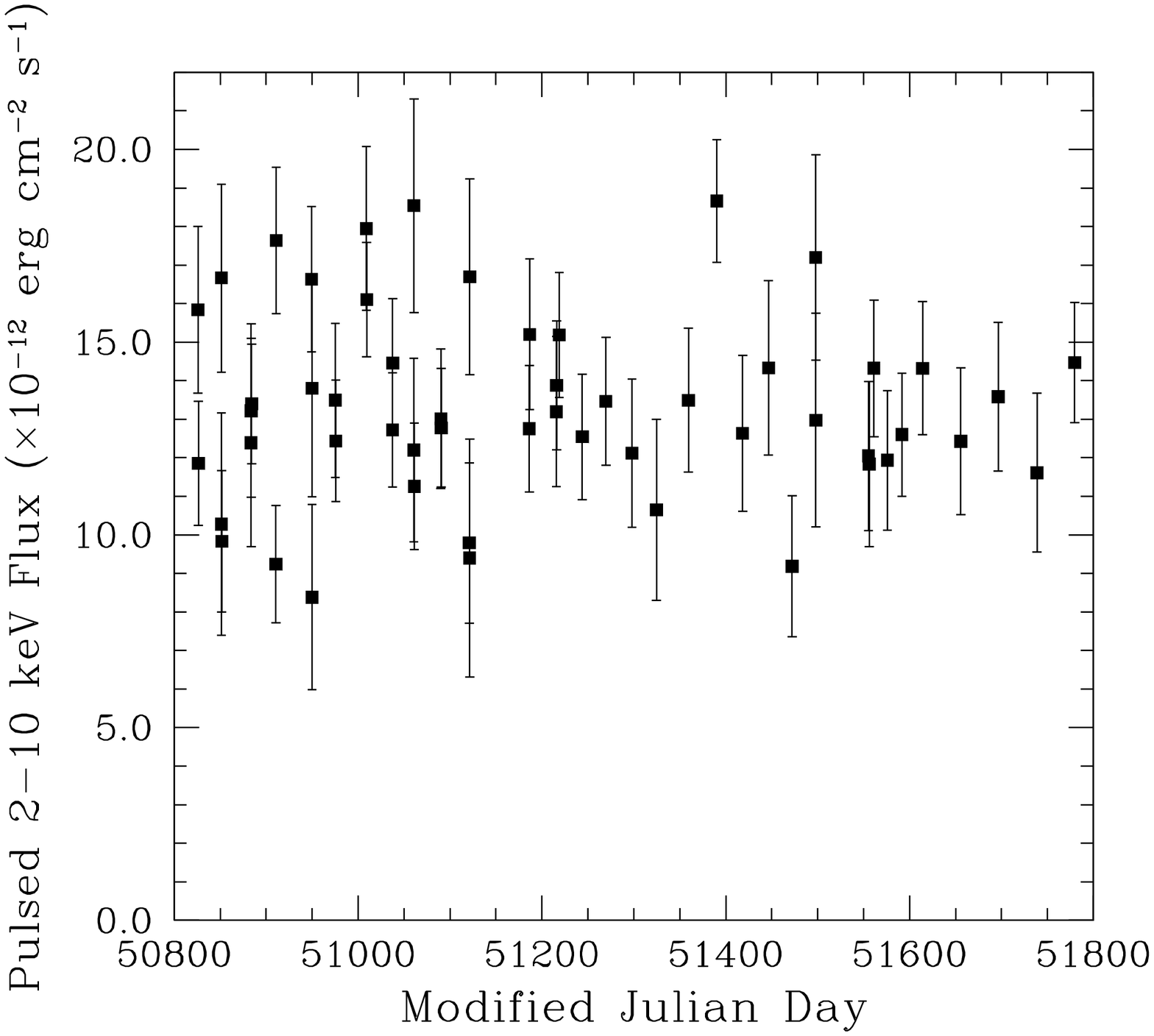}
\figcaption{Pulsed flux time series for \xte\ observations of \soe. Error bars represent $1\sigma$ confidence intervals. See \S\ref{sec:flux} for details of the analysis procedure.
\label{fig:flux1708}}
\end{figure}

\begin{figure}
\plotone{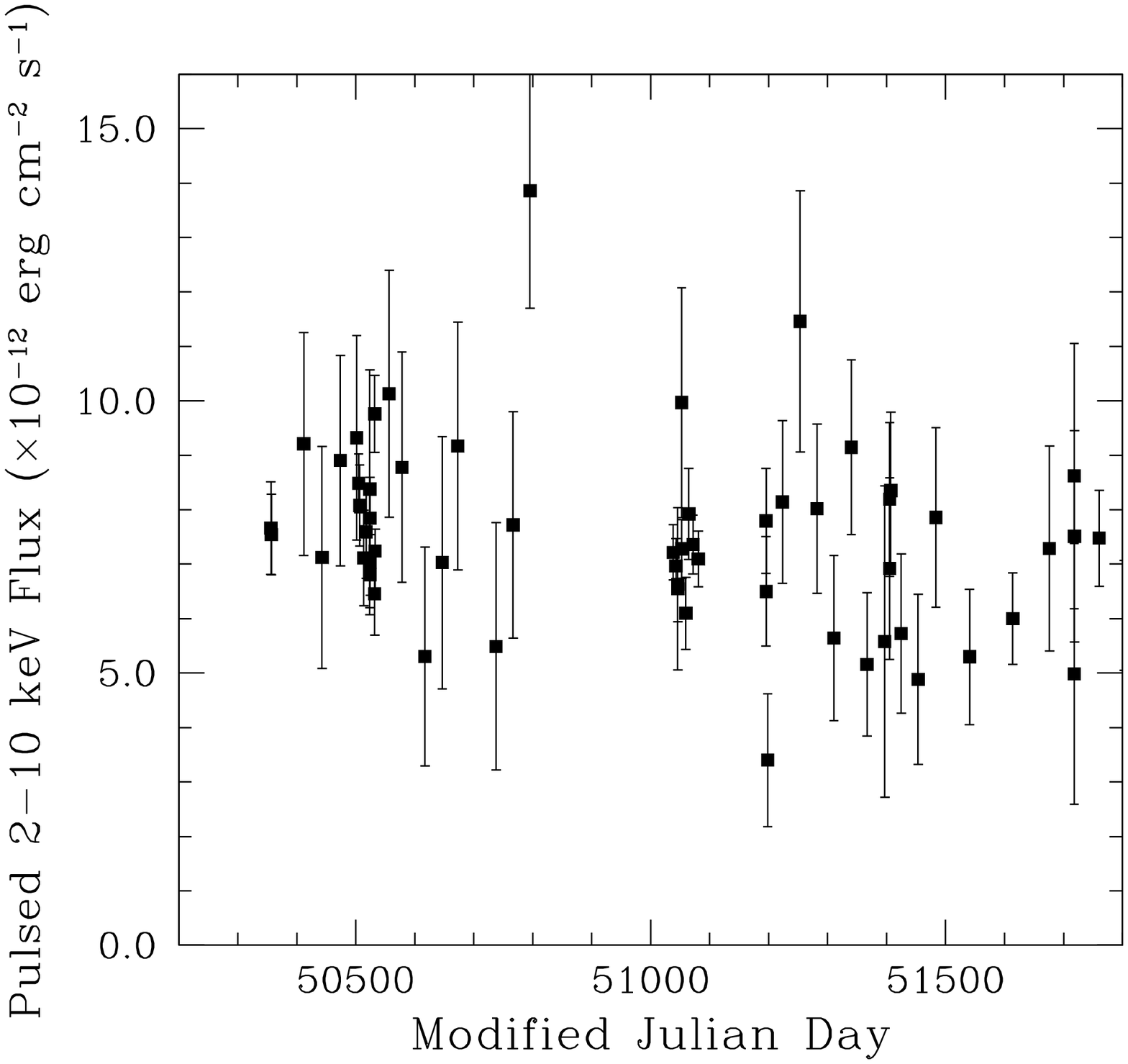}
\figcaption{Pulsed flux time series for \xte\ observations of \tfn. Error bars represent $1\sigma$ confidence intervals. See \S\ref{sec:flux} for details of the analysis procedure.
\label{fig:flux2259}}
\end{figure}

Data from each observing epoch were folded at the expected pulse period as was done for the timing analysis. However, for the flux analysis, 16 phase bins were used across the pulse. For each phase bin, we maintained a spectral resolution of 128 bins over the PCA range. Given the broad morphologies of the average pulse profile, only one phase bin could be used as a background estimator. The pulse profiles were then phase aligned, so that the same off-pulse bin was used for background in every case. The remaining phase bins were summed, and their spectral bins regrouped using the \texttt{FTOOL} \texttt{grppha}, such that no bin had fewer than 20 counts after background subtraction. Energies below 2~keV and above 10~keV were ignored, leaving 17--21 spectral channels for fitting, depending on the data set. The regrouped, phase-summed data sets, along with the background measurement, were used as input to the X-ray spectral fitting software package \texttt{XSPEC}\footnote{http://xspec.gsfc.nasa.gov}. Not all PCUs were on during our observations, but this was taken into account when producing response matrices. Response matrices were created using the \texttt{FTOOL}s \texttt{xtefilt} and \texttt{pcarsp}. Because of the limited spectral resolution, fitting a two-component model was not practical, so we used a simple photoelectrically absorbed power-law, at first holding only $N_H$ fixed (see Table~\ref{ta:flux}).  For all sources we found that the photon index $\Gamma$ was constant within the uncertainties; we therefore held it fixed at its mean value.  To extract a pulsed flux at each observing epoch we refitted each spectrum by varying only the normalization. Uncertainties were  measured using the \texttt{XSPEC} command \texttt{steppar}. Our flux time series for \oft, \soe\ and \tfn\ are displayed in Figures~\ref{fig:flux0142}, \ref{fig:flux1708} and \ref{fig:flux2259} respectively. We do not find evidence for any large variability in the pulsed flux. Although the measured $\chi_{\nu}^2$ statistics for \tfn\ and \soe\ are significantly larger than unity, given that the error bars are statistical only, we cannot rule out unmodeled systematic errors.  A long-term flux monitoring campaign with an imaging telescope is obviously desirable.

\section{Discussion}
\label{sec:discussion} 

\subsection{Timing}
\label{sec:disc_timing}

With our report of successful phase-coherent timing for \oft, the
detailed long-term timing properties of all known AXPs are becoming
clear.  As a population, they are generally spinning down steadily and
with impressive stability.  Indeed, for \tfn, the stability is
comparable to that of many radio pulsars: the $\Delta_8$ statistic
(defined as the phase deviation due to $\ddot{\nu}$ after $10^8$~s;
see \citeauthor{antt94} \citeyear{antt94}) is 0.43, consistent with
what would be expected for a radio pulsar having a comparable spin-down
rate, $-$0.78, given the large scatter in the radio pulsar values.  
For \oft, fitting for $\ddot{\nu}$ in either of the two best solutions
yields $|\ddot{\nu}| = (5 \pm 1) \times 10^{-24}$~s$^{-3}$, implying
$\Delta_8 = 0.89$, slightly larger than that expected for a radio
pulsar with the same spin-down rate, $-0.42$, though still comparable given
the scatter. 
For \soe, $\Delta_8$ is significantly higher than would be
predicted (2 versus 0.2), however this is consistent with glitch recovery.

The stability  of \tfn\ is particularly surprising given past reported
large deviations from simple spin-down \citep[e.g.][]{bs96}.  This is
in contrast to \tfe\ which also showed apparent deviations from simple
spin-down \citep[see, e.g.][]{opmi98,pkdn00} but which we have verified
in \xte\ monitoring \citep{kgc+01}.  As argued by \citet{bss+00},
it may be a coincidence that \tfn\ has shown great rotational stability
during our observations only (but see also \S~\ref{sec:disc_flux}).  


The current stability of \tfn\ allows for a test of the \citet{mel99} model of
radiative precession.  In that model, given well-defined but uncertain
assumptions about the geometry and location of the magnetic field, a
highly magnetized neutron star could deviate from sphericity and
exhibit significant precession with period of a few years.  This was
suggested as an explanation for timing anomalies seen for \tfn\ and
\tfe.  The current data for \tfn\ rule out, with 99\% confidence,
an amplitude for precession 0.013 times that used by \citet{mel99} to
explain the timing anomalies.  Radiative precession was also
ruled out by \xte\ monitoring for \tfe\ \citep{kgc+01}.  Frequent
large glitches in both these sources, as were suggested by
\citet{hh99}, also have not been confirmed.   Also, neither
\oft\ and \efo\ \citep{ggk+01} show any evidence for either radiative precession
or frequent large glitches.

Overall, however, a striking property of the timing of the AXPs as a
whole, apart from the stability, is the diversity of behavior.
\tfn\ and \oft\  have shown great stability;
\efo\ shows stability but with considerably more
red noise \citep{ggk+01}; \soe\ has exhibited a glitch with recovery
similar to what is observed for radio pulsars; and \tfe\ is less stable
than even some known accretors \citep{kgc+01}.

Interestingly, there is a correlation between timing stability and
$\dot{\nu}$ in our data.  The sources with by far the smallest
$\dot{\nu}$'s, \tfn\ and \oft, are the most stable (at least during our
observations), while that with the largest, \tfe, is the least so.  The
SGRs, being even less stable \citep{wkf+00,wkp+01}, have even larger
$\dot{\nu}$'s, in agreement with this trend.  \efo\ and \soe\ have
timing stabilities and $\dot{\nu}$ intermediate between the extremes.
One therefore might expect \efo\ to exhibit a glitch in the near
future.  If correct, the trend suggests a continuum of timing
properties between the AXP and SGR populations, lending additional
support to the connection between them.  Furthermore, it has long been 
recognized that radio pulsar timing stability is also correlated with spin-down rate
\citep[e.g.][]{ch80,antt94}.  This suggests, in addition, a connection
between all three populations, and a continuum of timing behavior that
depends on stellar magnetic field.  If correct, this provides strong
support for the magnetar model.

\citet{mw01} noted a correlation between spin-down rate and spectral properties
in AXPs.  An additional correlation with timing stability could strengthen
the argument that the spin properties, emission mechanisms, and internal
structure and dynamics are all related in these objects.  

%

\subsection{Pulse Morphology}

\begin{figure}
\plotone{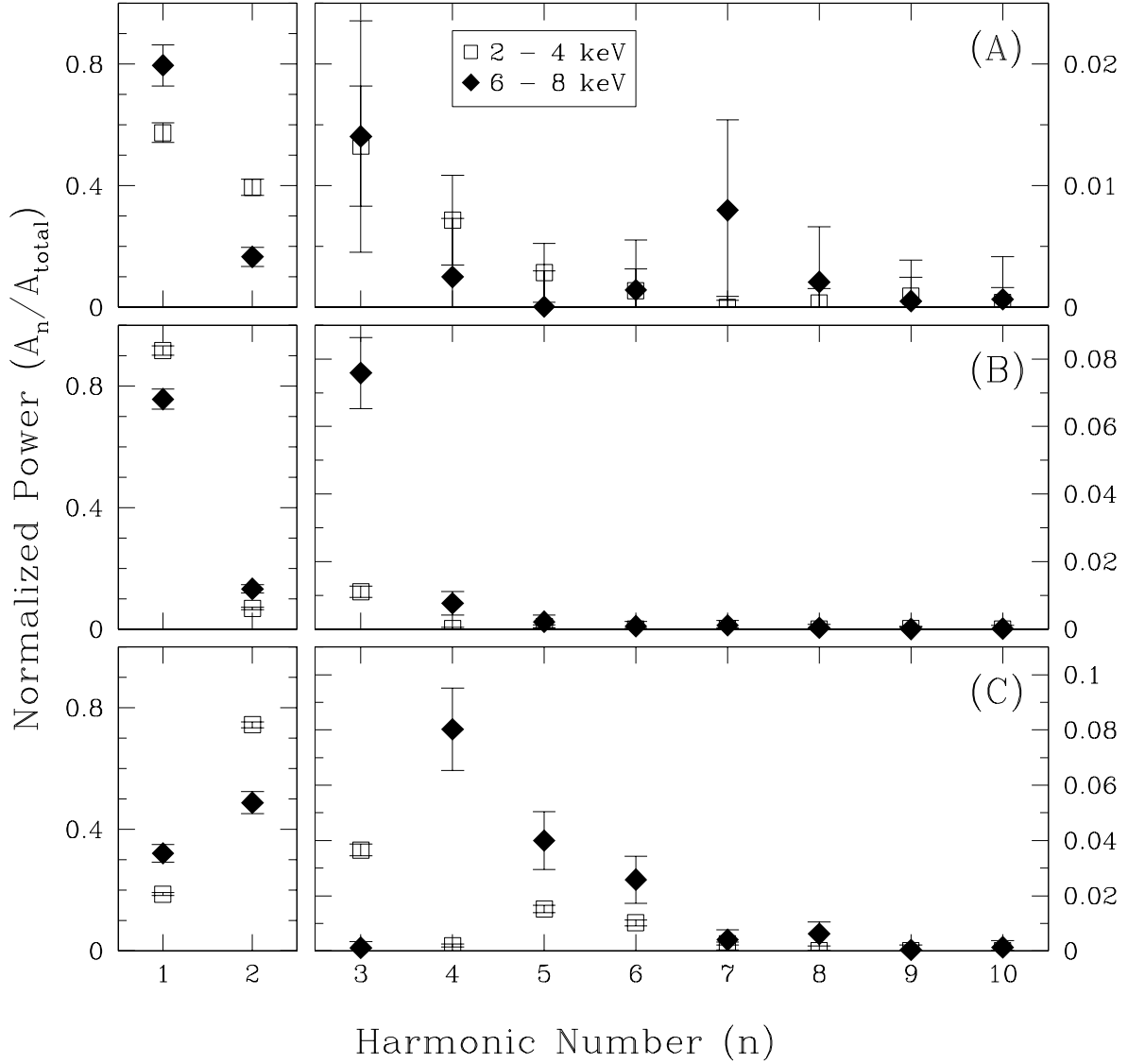}
\figcaption{Harmonic content of the average pulse profiles in Figure~\ref{fig:energy} in three energy bands. (A)~\oft; (B)~\soe; (C)~\tfn. The ratio of the power in the $n^{\textrm{th}}$ harmonic to the total power in all harmonics is plotted versus  $n$.
\label{fig:fft_energy}}
\end{figure}

The presence of substantial harmonic content in AXP pulse profiles (see
Fig.~\ref{fig:fft_energy}) is at odds with models in which the X-ray
emission is thermal from the surface of a hot neutron star having
conventional magnetic field.  As shown by \citet{dpn01}, the expected range of
ratios of second to first Fourier amplitudes, $A_2/A_1$, can be predicted for a
neutron star emitting thermal X-rays, as a function of total pulse
fraction, for different radially peaked beaming functions, surface hot-spot geometries,
and neutron star compactnesses.  Low harmonic content is expected primarily
because of gravitational light bending, unless there is significant
beaming.  Given the pulsed fraction of \tfn\ \citep[][and references
therein]{opk01}, \citet{dpn01} show that the ratio of $A_2/A_1 \leq 1$
for all possible geometries and compactnesses, while
Figure~\ref{fig:fft_energy}C shows $A_2/A_1 = 3.97 \pm 0.01 $ in the soft 2--4~keV band.  
The presence of a
significant third harmonic in \soe\ (Fig.~\ref{fig:fft_energy}B) and even a
significant sixth harmonic in \tfn\ (Fig.~\ref{fig:fft_energy}C) is
similarly problematic.  On the other hand, \citet{oze01} suggests that,
in the magnetar model, when properly accounting for the propagation of
X-rays through the neutron star atmosphere, the pulse morphology can be
surprisingly non-sinusoidal.  In particular, in $10^{14}-10^{15}$~G
magnetic fields, the emission from a surface patch can be highly and
non-radially beamed.  Furthermore, if the surface distribution itself is
non-isotropic, AXP profiles may be reproduceable.  
Indeed, \"Ozel, Psaltis \& Kaspi (2001) \nocite{opk01}
show that AXP pulsed fractions and luminosities can only be reproduced,
assuming thermal emission from the surface of a magnetar propagating
through a hydrogen atmosphere, with a single hot spot.  In addition,
Thompson, Lyutikov \& Kulkarni (2001) 
\nocite{tlk01} argue that AXP pulse profiles can be accounted for even
if the surface emission were isotropic, because of resonant scattering
off charged particles in the magnetosphere, which has anisotropic
optical depth.  A detailed comparison of model predictions with
observed AXP pulse profiles may prove interesting.

\subsection{Pulse Morphology Energy Dependence}

The energy dependence of the pulse profiles of AXPs can in principle be
a strong constraint on the emission model.  We have studied this by
comparing the morphologies of pulse profiles in two energy bands.  The
AXPs display a variety of behaviors in this regard.  For example, while
\tfe\ and \efo\ show essentially no change in pulse morphology with energy
\citep{kgc+01,ggk+01}, \soe\ shows tremendous variation
(Fig.~\ref{fig:energy}B).  In no case (except possibly \soe) are the
profiles in the two energy bands very different, problematic for interpreting
AXP two-component spectra as being a result of two independent emission
mechanisms \citep[e.g.][]{phh+01}. Figure~\ref{fig:energy}A shows a possible energy-dependent phase lag, in which the low energy pulse lags the high energy pulse, however, this is not observed in any other AXP.

There is a possible rough trend in which sources with the least
harmonic content (\tfe, \efo) show the least energy dependence.
However the lack of strong pulse morphology evolution with energy in
\tfn\ compared with \soe\ does not obviously support this.  Another possibly
interesting trend is that for the sources in which the ratio of the
Fourier amplitudes of the second and first harmonics is large (\tfn,
\oft), the ratio is largest at low energies.  This manifests itself in
the profile plots as the ratio of peak heights being closer to unity in
the hard band than in the soft, which is true of \soe\ as well.

Whether and how these observations can constrain models remains to be seen.
\citet{oze01} shows that, when photon propagation through a magnetar
atmosphere is properly modeled, the angular dependence of the
emission has two components: a narrow pencil beam at small angles with respect
to the surface normal, and a broad fan beam at large angles.  She shows
that the importance and opening angle of the pencil beam increases
with increasing photon energy.  This may be able explain the observed change
in the relative peak amplitudes as a function of photon energy,
especially in \soe, if part of the pulse is from a fan beam and part
from a pencil beam.  However detailed pulse profile modeling is required.

\subsection{Pulsed Flux Time Series}
\label{sec:disc_flux}

The pulsed fluxes of all known AXPs are roughly constant, with RMS fluctuations 
of at most $\sim$20--30\% of the mean ($1\sigma$) (see Table~\ref{ta:flux},
\citeauthor{kgc+01} \citeyear{kgc+01}, \citeauthor{ggk+01}
\citeyear{ggk+01}).  Past studies of the flux stability of two AXPs,
\tfn\ and \tfe, have suggested that their phase-averaged fluxes are
highly variable, with fluctuations possibly as large as a factor of ten
\citep{bs96,opmi98}.  Large flux variations, particularly on time
scales under a year and in the absence of torque variations, are
unexpected in the magnetar model.  This is because in this model,
X-rays come from the hot surface where the heat is maintained by the
decay of the interior magnetic field.  As the thermal conduction time
from core to surface is roughly a year \citep{rem91}, flux variations
on shorter time scales are hard to explain in the absence of burst activity 
as is seen in the SGRs \citep[e.g.][]{wkp+01}.

With the caveat that we measure pulsed flux while other studies
measured total flux, it is surprising that we find such flux stability
in our observations given past claims.  \citet{bss+00} 
argued that the flux stability of \tfn\ is a result of its
currently being in a quiescent accreting state that is also
characterized by timing stability (see also \S~\ref{sec:disc_timing}).
Though we cannot rule this out, we point out
that (i) the same coincidence would have to be true of \tfe\ for flux
stability, (ii) the pulsed fluxes of \oft, \soe\ and \efo\ are also
stable, and (iii) the pulsed flux of \tfe\ is stable even though it
shows unstable timing behavior, which argues that the flux stability is
independent of spin-down behavior.  We note that our flux measurements,
in contrast to past studies, have been made using a single
instrument, single bandpass,  and analysis method, which eliminates the 
difficulty in comparing measurements made with different instruments, 
different spectral ranges,  and different analysis methods.

\citet{vgtg00} have suggested that the variable X-ray pulsar
\axj\ which is at the center of the supernova remnant G29.6+0.1 may
also be an AXP.  This source has shown variations in its X-ray
luminosity by a factor of $\sim$10 between observations taken 6~yr
apart.  Given that the pulsed fluxes of all known AXPs are stable, the
interpretation of \axj\ as a {\it bona fide} AXP is therefore questionable.

%

%
%
%

\section{Conclusions}
\label{sec:conclusions}

We have presented a variety of observational results for three anomalous
X-ray pulsars, \oft, \soe\ and \tfn, obtained using regular monitoring
observations and archival observations from the PCA aboard {\it RXTE}.
Our results, combined with those for \tfe\ and \efo\ \citep{kgc+01,ggk+01}, 
provide significantly improved descriptions of AXP X-ray
properties that can be used to test models for the nature of these sources. 

Specifically, with our successful phase-coherent timing of \oft, the
detailed timing properties of all known AXPs are now established and
can be compared.  We find a wide variety of timing behaviors, ranging
from high stability (in \tfn\ and \oft), to instabilities so severe
that phase-coherent timing is not possible (in \tfe).  We note that
timing stability appears correlated with decreasing $\dot{\nu}$.  If
this trend is real, it suggests a continuum of timing properties
between the AXP and SGR populations, lending additional support to the
connection between them.  We note that a correlation between timing
instability and spin-down rate has long been recognized among radio
pulsars \citep[e.g.][]{ch80,antt94}, suggesting a connection between
all three populations.

We have also used the {\it RXTE} data to investigate a variety of other
AXP properties.  Motivated by a report of a significant change in pulse
morphology in \tfn\ in 1990 \citep{ikh92}, we searched for changes in
the pulse morphology at each epoch for all targets.  We find no
significant pulse morphology variations.  Typically, we rule out variations
in features having amplitude $\gapp$20\% of the peak amplitude at the
$1\sigma$ level, although the limit depends on source and integration time.

We have presented high signal-to-noise average pulse profiles for
each AXP, and considered them as a function of energy.  We show that, as
in the timing properties, there is a variety of different behaviors for
the energy dependence.  Possible trends include a greater energy
dependence for pulses profiles having greater harmonic content, and,
for the latter sources, relative peak amplitudes tending
closer to unity as photon energy increases.  Detailed modeling of
AXP pulse profiles, in addition to their pulsed fractions and
luminosities \citep{opk01} can test models in which the emission is from
the surface of a cooling magnetar \citep{oze01,hl01,ztst01}.

Finally, we use the monitoring and archival data to obtain pulsed flux
time series for each source.  We have found no large changes in pulsed
flux for any source, and have set $1\sigma$ upper limits on variations
$\sim$20--30\% (depending on the source).  This is surprising given
previous reports of large (factor of 5--10) total flux variations in
\tfn\ and \tfe\ \citep{bs96,opmi98}.  Assuming a constant pulsed
fraction, this suggests that  more than one of the AXPs happen to be much more
quiescent during the {\it RXTE} monitoring than in the past.

\acknowledgments

We are grateful to D.~Chakrabarty, M.~Lyutikov, M.~Muno, F.~\"Ozel, D.~Psaltis, M.S.E.~Roberts and C.~Thompson for useful discussions.   We also thank B. Paul and F. Nagase for
assistance in obtaining the {\it Ginga} data.  This work was supported in part by a NASA LTSA grant (NAG5-8063) and an NSERC Research Grant (RGPIN228738-00) to VMK, with additional support from a NASA ADP grant (NAG 5-9164). This research has made use of data obtained through the High Energy Astrophysics Science Archive Research Center Online Service, provided by the NASA/Goddard Space Flight Center.


\begin{deluxetable}{lcccc}
\tablecolumns{5}
\tablewidth{400pt}
\tablecaption{Summary of \xte\ Observations \label{ta:exposures}}
\tablehead{
\multicolumn{1}{l}{Observing Period} & \colhead{Nominal} & \colhead{Nominal} &\colhead{Number} &\colhead{Total} \\
\colhead{} & \colhead{Exposure} & \colhead{Separation} &\colhead{of Obs.} &\colhead{Exposure} \\
 \colhead{}&\colhead{(ksec)}& \colhead{(weeks)}   & \colhead{} &\colhead{(ksec)} }
\startdata
\cutinhead{\oft}   
	 Mar 1996             & \nodata & \nodata & 1  & 7.0   \\
	 Mar 1996             & \nodata & \nodata & 1  & 43.5   \\
         Mar 1998             & \nodata & \nodata & 1  & 20.3   \\ 
	 Nov 1996 -- Dec 1997 & 1.1      & 4--5     & 14 & 14.9   \\
	 Mar 2000 -- Feb 2001 & 3.3      & 4--5     & 14 & 46.0   \\
	 Mar 2001 -- Jun 2001 & 6.6      & 4--5     & 3  & 19.9   \\
\cutinhead{\soe}    
         Jan 1998 -- Jan 1999 & 6.2      & 4--5      & 12 & 74.0   \\
	 Feb 1999 -- Jun 2001 & 3.1      & 4--5      & 33 & 103.6  \\
	 May 2001             & 2.0      & 1/7      & 5  & 10.1   \\
\cutinhead{\tfn}      
	 Sep 1996             & \nodata & \nodata & 1  & 86.9   \\
         Nov 1996 -- Dec 1997 & 1.0      & 4--5     & 14 & 14.0   \\
	 Feb 1997 -- Mar 1997 & 24.8     & 1        & 5  & 124.1  \\
	 Aug 1998 -- Sep 1998 & 14.0     & 1        & 8  & 111.7  \\
	 Jan 1999 -- Mar 2001 & 3.2      & 4--5     & 19 & 61.3   \\
	 Mar 2000 -- May 2001 & 7.0      & 4--5     & 11 & 77.3   \\
\enddata
\end{deluxetable}

\clearpage
\begin{deluxetable}{lcccc}
\tablecaption{Spin Parameters for \oft.\tablenotemark{a} \label{ta:timing0142} }
\tablehead{
\colhead{} & \colhead{First Span} & \colhead{Second Span}  & \colhead{A} & \colhead{B} }
\startdata
MJD Range         & 50411--50893  & 51610--52028    & 50411--52028   & 50411--52028       \\
No. TOA           & 15            & 17              & 32             & 32                 \\
$\nu$ (Hz)        & 0.115096877(10)  & 0.1150969336(5) & 0.1150969299(10) & 0.1150969209(11)   \\
$\dot{\nu}$ ($10^{-14}$ Hz/s) & $-$2.687(5)    & $-$2.649(11)    &   $-$2.5980(23)  & $-$2.5969(24)  \\
 Epoch (MJD)      & 51704.000     & 51704.000       & 51704.000     & 51704.000         \\
RMS residual   & 0.027  & 0.011           & 0.074         & 0.077             \\
\enddata
\tablenotetext{a}{Ephemerides A and B are both viable given the data set, which includes a 2-yr gap at the center of the 4.4-yr span.  Ephemeris B has one additional pulse in the gap relative to Ephemeris A.  See \S~\ref{sec:timing}.}
\end{deluxetable}

\clearpage
\begin{deluxetable}{lcc}
\tablewidth{350pt}
\tablecaption{Spin Parameters for \soe\ and \tfn. \label{ta:timing} }
\tablehead{
\colhead{} & \colhead{\soe}  & \colhead{\tfn} }
\startdata
MJD Range                         & 51472--51995   & 50356--52016       \\
No. TOA                           & 19             & 67                 \\
$\nu$ (Hz)                        & 0.090917063(5) & 0.14328806234(8)   \\
$\dot{\nu}$ ($10^{-14}$ Hz/s)     & $-$16.07(2)    & $-$0.99434(16)     \\
$\ddot{\nu}$ ($10^{-23}$ Hz/s$^2$)& 5.2(6)         & 0.228(14)          \\
 Epoch (MJD)                      & 51215.9308     & 51995.5827         \\
RMS residual (periods)            & 0.0075         & 0.0099             \\
\enddata
\end{deluxetable}

\clearpage
\begin{deluxetable}{lccccccc}
\tablecolumns{8}
\tablewidth{425pt}
\tablecaption{AXP Spectral Parameters and 2-10~keV Pulsed Flux Variability Upper Limits  \label{ta:flux}}
\tablehead{
\colhead{} & \multicolumn{3}{c}{Spectral Parameters} &\colhead{}& \multicolumn{3}{c}{Flux Time Series Parameters} \\
\cline{2-4} \cline{6-8} \\
\multicolumn{1}{l}{Source}                            & \colhead{$\log N_H$ \tablenotemark{a} }                      &  
\colhead{Refs. \tablenotemark{b} }          & \colhead{$\Gamma$ \tablenotemark{c} }                       &\colhead{} & 
\colhead{$\chi_{\nu}^2$ \tablenotemark{d} } & \colhead{Degrees of} & \colhead{$\unitfrac{RMS flux}{mean flux}$ \tablenotemark{e}} \\
\colhead{}                                  & \colhead{(cm$^{-2}$)}                                        & 
\colhead{}                              &\colhead{}    &\colhead{} & \colhead{}                                                   & 
\colhead{Freedom}                                  & \colhead{}                                                   }

\startdata

\oft    & 22.18& 1 &3.50 && 0.63 & 36& 0.26 \\    
\soe    & 22.25 &2& 2.63 && 1.45 & 52&0.17 \\            
\tfn    & 22.08 &3& 4.02 && 1.38 & 59&0.28 \\

\enddata
\tablenotetext{a}{For each AXP, $N_H$  was held fixed at these values.}
\tablenotetext{b}{(1)~\citet{ms95};  (2)~\citet{snt+97}; (3)~\citet{bs96}.}
\tablenotetext{c}{Mean power-law index as measured for the pulsed portion of the spectrum.}
\tablenotetext{d}{$\chi_{\nu}^2$ statistic of the pulsed flux time series displayed in Figures~\ref{fig:flux0142},~\ref{fig:flux1708} and \ref{fig:flux2259}.}
\tablenotetext{e}{RMS of the pulsed flux time series displayed in  Figures~\ref{fig:flux0142},~\ref{fig:flux1708} and \ref{fig:flux2259} in units of the mean pulsed flux.}
\end{deluxetable}

\clearpage

\clearpage

\clearpage

\clearpage

\clearpage

\clearpage
\begin{figure}
\plotone{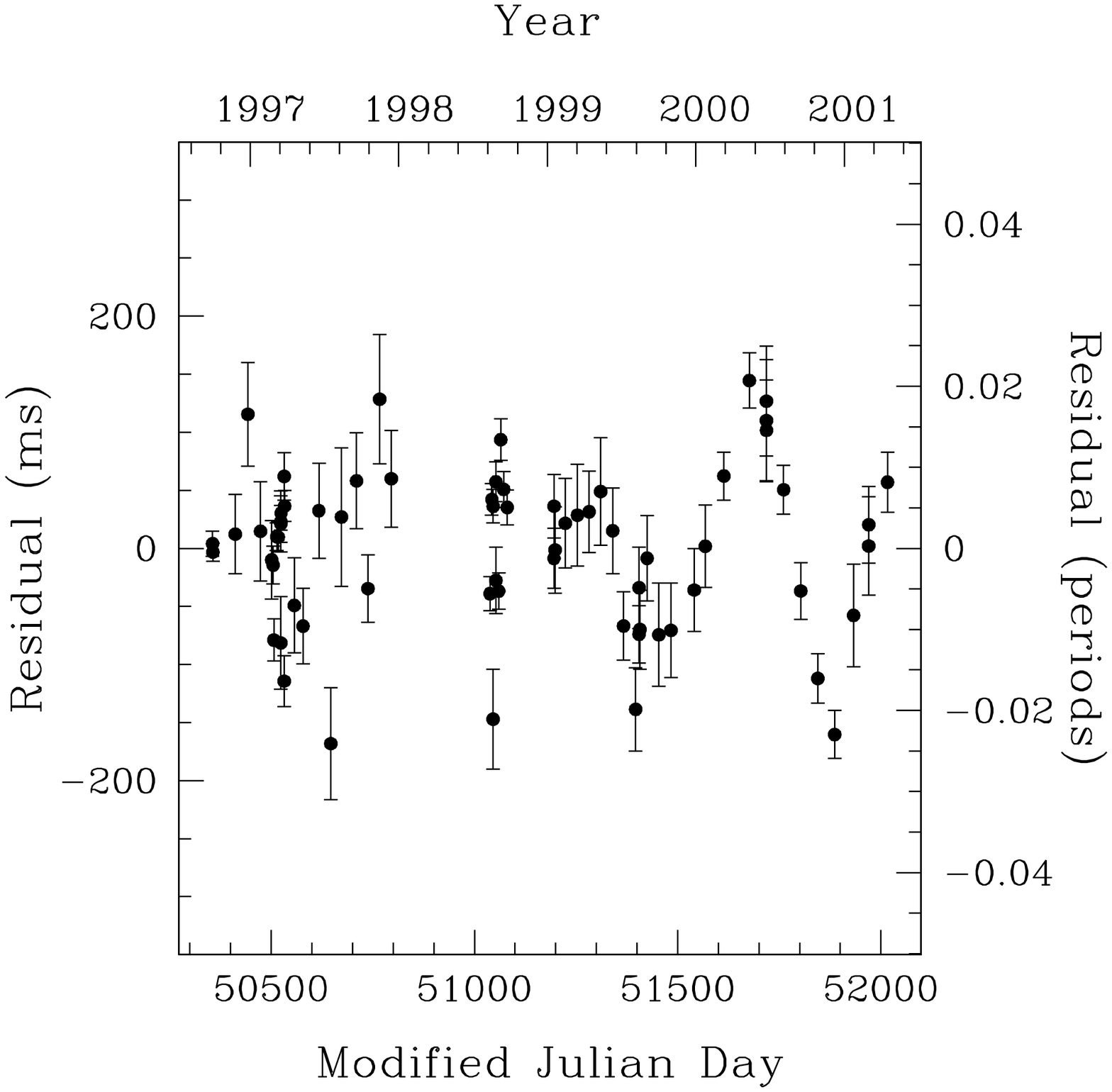}
\figcaption{Arrival time residuals for \tfn\ with $\nu$, $\dot{\nu}$ and $\ddot{\nu}$ subtracted.
\label{fig:res2259}}
\end{figure}

\clearpage

\clearpage

\clearpage

\clearpage

\clearpage

\end{document}